# Deterministic Role of Concentration Surplus of Cation Vacancy over Anion Vacancy in Bipolar Memristive NiO

*Zhong Sun,* [†,⊥] *Yonggang Zhao,* [*,†,⊥] *Min He,* [∥] *Lin Gu,* [∥,⊥] *Chao Ma,* [∥] *Kuijuan Jin,* [∥,⊥] *Diyang Zhao,* [†,⊥] *Nannan Luo,* [†,⊥] *Qinghua Zhang,* [#] *Na Wang,* [†,⊥] *Wenhui Duan,* [†,⊥] *Ce-Wen Nan* [#]

[†]Department of Physics and State Key Laboratory of Low-Dimensional Quantum Physics, Tsinghua University, Beijing 100084, China

[∥]Beijing National Laboratory for Condensed Matter Physics, Chinese Academy of Sciences, Beijing 100190, China

[⊥]Collaborative Innovation Center of Quantum Matter, Beijing 100084, China

[#]State Key Lab of New Ceramics and Fine Processing, School of Materials Science and Engineering, Tsinghua University, Beijing 100084, China

**ABSTRACT:** Migration of oxygen vacancies has been proposed to play an important role in the bipolar memristive behaviors since oxygen vacancies can directly determine the local conductivity in many systems. However, a recent theoretical work demonstrated that both migration of oxygen vacancies and coexistence of cation and anion vacancies are crucial to the occurrence of bipolar memristive switching, normally observed in the small-sized NiO. So far, experimental work addressing this issue is still lacking. In this work, with conductive atomic force microscope and combined scanning transmission electron microscopy & electron energy loss spectroscopy, we reveal that concentration surplus of Ni vacancy over O vacancy determines the bipolar memristive switching of NiO films. Our work supports the dual-defects-based model, which is of fundamental importance for understanding the memristor mechanisms beyond the well-established oxygen-vacancy-based model. Moreover, this work provides a methodology to investigate the effect of dual defects on memristive behaviors.

## 1. INTRODUCTION

Memristive devices (or memristors) are resistive switching devices[1] that have potential applications in nonvolatile memory (resistive random access memory, abbreviated RRAM),[2] logic circuits,[3,4] neuromorphic networks,[5,6] *etc.* Applications of memristors call for vigorous investigation on the mechanisms, which is essential for reliability, scalability and optimization of such applications.[4,7,8] So far, most mechanisms of resistive switching involve ionic migration,[1,2,4] and can be categorized in general into three types,[2] including electrochemical metallization mechanism (ECM), valence change mechanism (VCM) and thermochemical mechanism (TCM). Migration of anion (usually oxygen ion, or equivalently its counterpart oxygen vacancy) plays an important role in both VCM and TCM, which show bipolar and unipolar resistive switching, respectively. NiO is an archetypical TCM material,[2] and the mechanism of its unipolar switching has been studied extensively, which is recognized as oxygen-vacancy-migration induced Ni filament connection/disconnection.[9-11] Most previous reports on NiO have been focused on the unipolar switching,[2] which is usually measured with pad-type electrodes. Interestingly, it has been found that when the device scales down, the resistive switching behaviors becomes bipolar,[12-16] suggesting that different mechanism is involved on the nanoscale. To understand its mechanism is important in terms of physics, as well as applications since devices are usually and increasingly miniaturized.

The bipolar resistive switching of NiO had been interpreted as a result of oxygen-vacancy-migration induced stoichiometry change,[14-16] resembling the oxygen-vacancy-based model of VCM in the *n*-type oxides.[2,17-19] However, the charge carrier type may has a decisive impact on the switching mechanism, as observed in the case of unipolar resistive switching of $TiO_2$ and NiO.[20-22] In NiO, oxygen vacancies have deeply localized defect levels and thus do not contribute to the electrical transport,[23,24] in contrast to the case in the *n*-type oxides, in which oxygen vacancies act as electron donors with defect levels locating near the conduction band bottom,[25,26] and their accumulation/depletion directly brings about conductivity alteration, resulting in memristive behaviors. Actually, the electrical transport in NiO is dominated by hole carriers, generated by nickel vacancies with shallow defect levels,[23,24] but nickel vacancies are considered more immobile than oxygen vacancies.[27-29] So the bipolar memristive behaviors in NiO cannot be explained by the aforementioned model. Of late, with first-principles calculations, Oka *et al.*[29] got the band structures of NiO with isolated nickel vacancy, isolated oxygen vacancy and paired nickel and oxygen vacancies, respectively. They suggested that it is the oxygen-ion-migration-induced concentration alteration of isolated nickel vacancies that causes the bipolar memristive behaviors of NiO. Up to now, there has been no experimental work to address this dual-defects-based model, probably due to the difficulty in characterizing both types of vacancies, especially oxygen vacancies.

Recently-developed aberration-corrected annular-bright-field (ABF) scanning transmission electron microscopy (STEM) has demonstrated its power in identifying individual light atoms such as oxygen,[30-32] enabling detecting the presence of oxygen vacancies, which is very difficult to achieve by other methods. Coupled with TEM, electron energy loss spectroscopy (EELS) is a technique broadly used to examine the chemical composition and bonding states of matter,[33,34] so it can be utilized to analyze the nickel vacancies in NiO when combined with the characterization of oxygen vacancies.

In this work, we investigated the dependence of bipolar memristive behaviors of NiO on both nickel and oxygen vacancies. The transport properties were measured with a conductive atomic force microscope (CAFM). By controlling the intrinsic oxygen content during sample preparation and extrinsic oxygen pressure during measurement, the bipolar memristive behaviors of NiO can be tailored. By using STEM and EELS, we have not only confirmed the coexistence of nickel and oxygen vacancies in NiO, but also revealed the deterministic role of concentration surplus of Ni vacancy over O vacancy in bipolar memristive behaviors of NiO. Our work provides an experimental support for the first time to the dual-defects-based model, which is of key importance for understanding the bipolar memristive behaviors of NiO, and is of vital significance for engineering further improvements. Moreover, we also provide a methodology to investigate the effect of dual defects on memristive behaviors.

## 2. EXPERIMENTAL SECTION

**2.1 Sample preparation.** NiO films were deposited on commercial Pt/Ti/SiO$_2$/Si substrates with a sintered ceramic NiO target by PLD, which was equipped with a KrF excimer laser ($\lambda$ = 248 nm, pulse energy = 310 mJ, frequency = 3 Hz, Lambda Physik). The deposition temperature was 500 ℃. According to the calibrated growth rate under different deposition oxygen pressures, we prepared NiO films with different oxygen contents and different film thicknesses.

**2.2 Electrical measurements.** The electric measurements were all accomplished by CAFM setups, including a Bruker Multi-mode V SPM with a CAFM modules, whose compliance current is 12.28 nA, and a Seiko SII E-SWEEP AFM with a current collector, whose compliance current is 100 pA, the former enables only measurement in atmosphere, while the latter can perform measurements under different ambient-oxygen pressures.

**2.3 Characterizations.** The crystal quality and orientation of NiO films were inspected on a Rigaku D/max-RB X-ray diffractometer with a Cu $K_\alpha$ radiation. The STEM images were obtained in an ARM－200CF transmission electron microscope operated at an acceleration voltage of 200 kV (JEOL, Tokyo, Japan) and equipped with double spherical aberration (Cs) correctors. The attainable resolution of the probe defined by the objective pre-field is 78 picometers. The TEM and EELS measurements were operated on a Tecnai F20 transmission electron microscope operated at an acceleration voltage of 200 kV (FEI, USA).

## 3. RESULTS AND DISCUSSION

NiO films were prepared by pulsed laser deposition (PLD) with Pt as bottom electrode (BE), and show fine columnar structures with a columnar grain size of about 20-30 nm, as shown by the AFM (Figure 1a) and STEM results (Figure S1). Electrical measurements were accomplished by CAFM, which was equipped with Pt/Ir-coated tips with a curvature radius of 20 nm. During the measurement, the tip was grounded while a bias voltage was applied on the Pt BE. The schematic sample layout and measurement configuration are displayed in Figure 1a. As shown in Figure 1b, the samples show typical bipolar resistive switching behavior. The voltage is swept from -10 V, across zero, to +10 V, resulting in the low resistance state (LRS), and this is the SET process. The following voltage-sweep from +10 V back to -10 V results in the high resistance state (HRS), and is termed RESET process. Besides the local current-voltage (*I-V*) characteristics, the CAFM current mapping also demonstrates the bipolar resistive switching of NiO. In the inset of Figure 1b, the bright square was written with +5 V, then the inner dark square was erased with -5 V, and the overall current distribution was mapped with a +0.2 V read voltage. The apparently uniform distribution of currents is due to the large amount of conducting filaments in the LRS, where a single filament cannot be distinguished on such a large scale. Notably, an initial forming process is not required here for resistive switching, which was also reported in a previous work studying the bipolar memristive NiO film with CAFM by Lee *et al*.[35] The forming-free characteristic of memristive device is a significant advantage for practical applications. We also examined other memristor characteristics of NiO films, including the voltage-sweep rate dependence of *I-V* hysteresis loop, and the continuously tunable conductivity.[18,36,37] In Figure 1c, the voltage-sweep rate is increased from 0.4 V/s to 40 V/s, and the hysteresis loop (which is in semi-log scale for clarification) shrinks gradually as the voltage-sweep rate increases. Specifically, the *I-V* curves share almost the identical LRS, but the currents of HRSs are distinct under different voltage-sweep rates, and this can be understood in the framework of aforementioned dual-defects-based model accounting for anion migration induced resistive switching. If the voltage-sweep rate is low, the negative voltage sweep will cause migration of more oxygen ions, generating more oxygen vacancies adjacent to nickel vacancies in NiO, resulting in a more insulating HRS. When the voltage-sweep rate is above 4 V/s, the currents of HRSs are no more distinct, which is likely due to the reason that oxygen ions can no longer respond to the high voltage-sweep rate. The LRSs, which is determined by the intrinsic concentration of isolated nickel vacancies, are same for all the voltage-sweep rates with voltage sweeping back from +10 V. The shrink of *I-V* hysteresis loops with increasing voltage-sweep rate is a typical characteristic of memristors.[18,36] We also carried out five consecutive sweeps from 0 V to 10 V and back to 0 V and the results are shown in Figure 1d. It can be seen that the conductivity of NiO films can be continuously tuned, wherein every forward sweep retraces the previous backward sweep at low voltages until the conductivity reaches saturation. In addition, we monitored the current on a local point with a +3 V bias, and observed that the conductivity increased continuously (Figure 1d inset). The continuously tunable conductivity suggests the accumulation effect of oxygen ion migration. Such a feature of memristors is sometimes called analog-type switching,[1] and is desirable in multiple states memory and other applications such as analog circuits and neuromorphic computing. For example, the permanent transition to a higher-conduction state in the inset of Figure 1d suggests that the NiO memristor can be used to mimick the long-term potentiation (LTP) mechanism of a biological synapse.[38] The saturation of tunable conductivity is determined by the intrinsic isolated nickel vacancies as in the case of voltage-sweep rate dependence of *I-V* hysteresis loop. With respect to the reliability of device applications, the reproducibility and endurance of NiO memristor were tested. The results show that the bipolar resistive switching of NiO can be cycled uniformly for at least 1000 times (Figure S2). Therefore, investigation of the NiO films by CAFM shows typical bipolar memristive behaviors with characteristics of memristor, and good device performance of resistive switching.

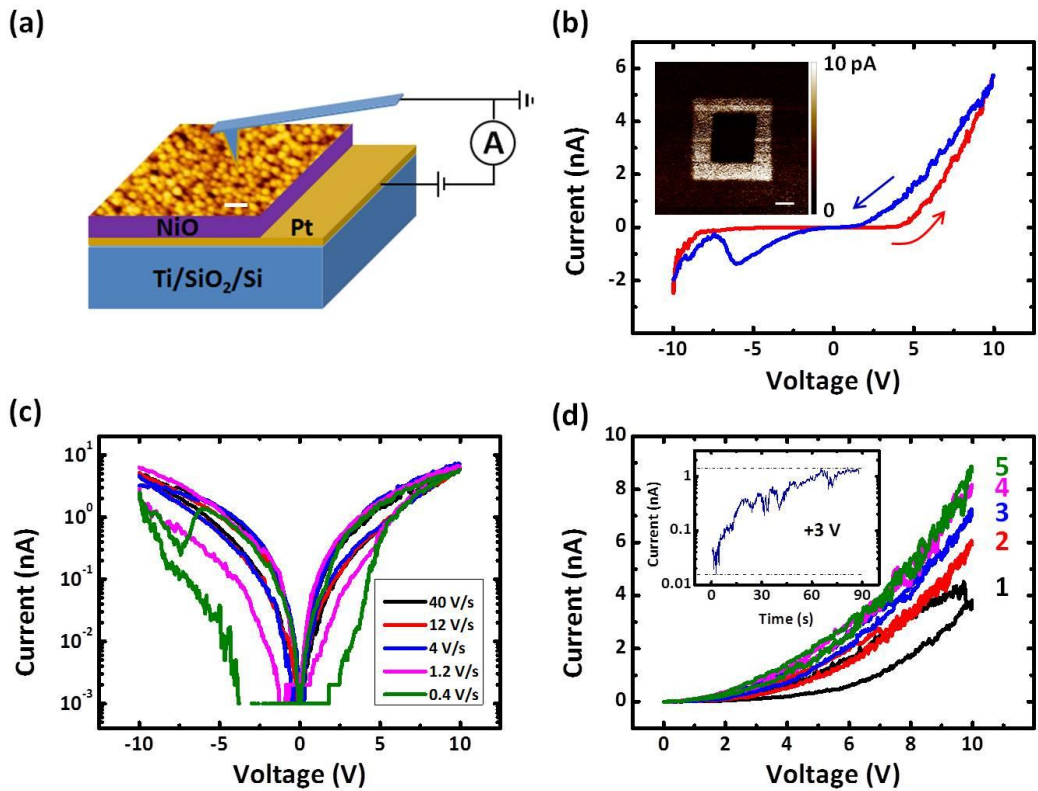

**Figure 1.** Memristor characteristics of NiO investigated with CAFM. (a) Schematic of sample layout and the measurement configuration with superposed AFM morphology of NiO film. Scale bar: 100 nm. (b) Typical bipolar resistive switching of NiO represented by *I-V* curve and CAFM

current mapping. Inset scale bar: 500 nm. (c) Voltage-sweep rate dependence of bipolar resistive switching. (d) Continuous conductivity enhancement under five consecutive positive voltage sweeps. The inset is the conductivity enhancement under a persistent +3 V bias.

We measured the memristive behaviors of NiO films with thicknesses of 30 nm, 50 nm and 100 nm, respectively, which were prepared by PLD at a 10 Pa oxygen pressure. The results in Figure 2a show that the magnitude of the resistive switching decreases as the film thickness increases. The 5-hours retention results of CAFM mapping show that the written LRSs are all stable in ambient conditions for these samples (Figure S3). Since both interfaces of tip/NiO and NiO/Pt-BE in this structure can form quasi-ohmic contact for hole conduction (Figure S4),[35] due to the high work functions of both the Pt/Ir coating of the CAFM tip and the Pt BE,[39,40] the thickness dependence of memristive behavior should be ascribed to its bulk effect nature, which describes that the oxygen-ion-migration alters the local conductivity via recombination of oxygen ions with oxygen vacancies and the remained isolated nickel vacancies constitute a *p*-type conducting filament throughout the NiO film during the SET process.[29] The conducting filaments have been observed in bipolar memristive NiO through CAFM mapping of LRS, in which filaments with diameters around 10 nm can be distinguished by tip applied with small contact force and resulting in contact radius of as tiny as 1.38 nm (Figure S5). When the film thickness increases, the distance that oxygen ions need to travel during the switching events becomes longer,[41] and the electric field inside the film becomes more diverging, which leads to lower travelling speed of oxygen ions, hindering occurrence of the bipolar resistive switching due to the absence of adequate stoichiometry change. Therefore, the thickness dependence of memristive behaviors reveals the bulk effect nature (which is filamentary type), consistent with the scenario depicted by the dual-defects-based model.[29]

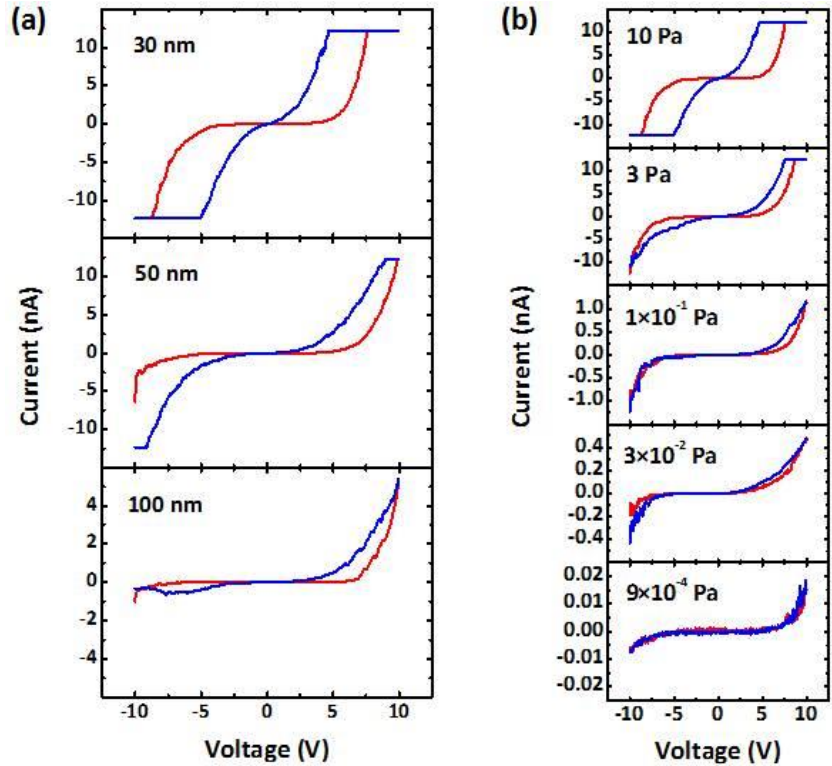

**Figure 2.** Film thickness and oxygen content dependences of bipolar resistive switching of NiO. (a) Bipolar resistive switching of 10 Pa-deposited NiO films with various thicknesses. (b) Bipolar resistive switching of 30 nm-thick NiO films deposited at various oxygen pressures.

In order to investigate the dependence of memristive behaviors on nickel and oxygen vacancies, we deposited 30 nm-thick NiO films at different oxygen pressures of 10, 3, $1\times10^{-1}$, $3\times10^{-2}$ and $9\times10^{-4}$ Pa, respectively, which would accordingly result in NiO samples with different concentrations of nickel and oxygen vacancies. It shows that the ratio of HRS/LRS decreases with decreasing deposition oxygen pressure (Figure 2b) and the resistive switching even disappears for the $9\times10^{-4}$ Pa-deposited NiO film. Similar behavior is also shown for the 100 nm-thick NiO films (Figure S6). We have also compared the resistive switching property of the as-grown sample deposited at low oxygen pressure and that annealed in oxygen for 30 min, and the results show that the resistive switching is facilitated for the annealed sample (Figure S7). Therefore, the bipolar memristive behaviors of NiO depends strongly on oxygen content. The difference between the sample grown at high oxygen pressure and that grown at low oxygen pressure will be revealed by STEM and EELS later.

Besides the inner-oxygen content controlled by deposition oxygen pressures and modulating the memristive behaviors intrinsically, the ambient oxygen can also be exploited as an extrinsic oxygen source to modulate the memristive behaviors of NiO, as the oxygen incorporation/extraction during resistive switching has been demonstrated for NiO film probed by CAFM.[14] As shown in Figure 3a-c, the contrast of CAFM current mapping between the LRS and HRS is apparent for measurement in air and almost disappears in current maps measured in vacuum, and appears again after the CAFM chamber being fulfilled with oxygen ($1 \times 10^4$ Pa). So, the ratio of HRS/LRS increases with ambient-oxygen pressure, similar to the dependence on deposition oxygen pressure. The ratios of HRS/LRS are presented in Supplementary Table S1. The *I-V* curves measured at different ambient-oxygen pressures conform to this phenomenon (Figure S8). On the other hand, we have demonstrated that moisture doesn't have notable impact on the bipolar memristive behaviors of NiO film (Figure S9), though it has been shown to play a significant role in the anionic memristive device.[42] Furthermore, we measured the ratios between the initial resistance state (IRS) and LRS at various ambient-oxygen pressures for the 10 Pa-deposited NiO film, and in atmosphere for NiO films deposited at various oxygen pressures, and the results are shown in Figure 3d. It shows that the comparable magnitude of resistive switching (the ratio of IRS/LRS) can be achieved by different combinations of intrinsic and extrinsic conditions (Figure 3e,f). Therefore, both the inner oxygen and the ambient oxygen play vital roles in determining the bipolar memristive behaviors of NiO, and they contribute collaboratively to the memristive behaviors with distinct contributions. As illustrated in the discussion later, the ambient oxygen is expected to modulate the local concentrations of nickel and oxygen vacancies near the surface of NiO films.

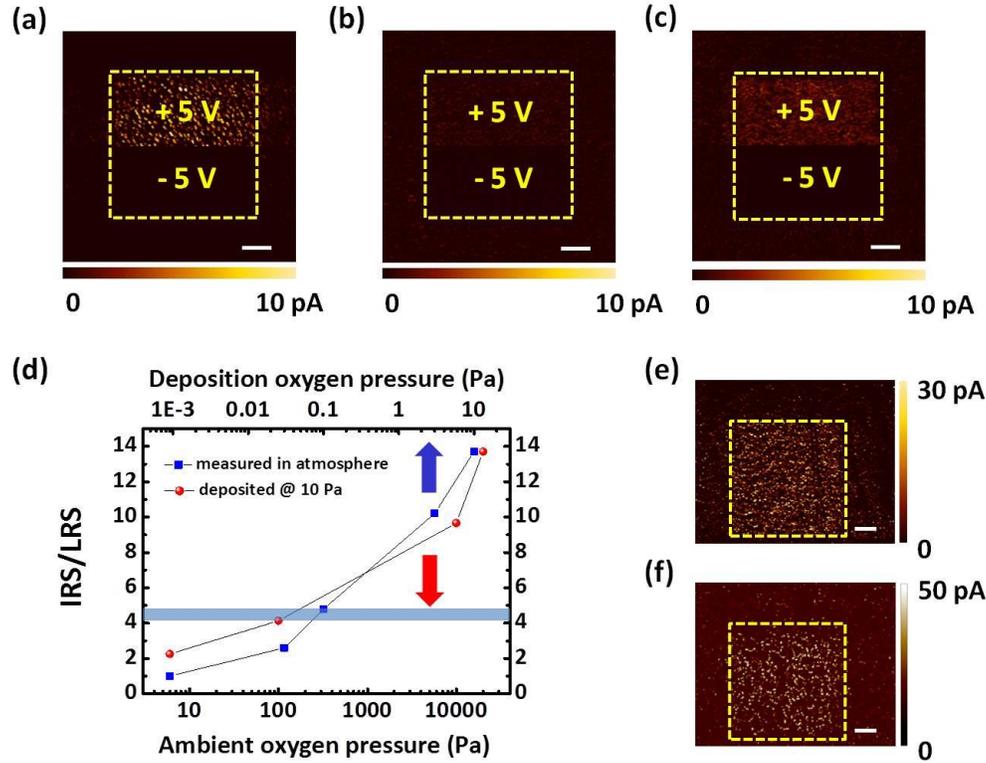

**Figure 3.** Collaborative effect of inner-oxygen content and ambient-oxygen pressure on resistive switching of NiO. CAFM current mappings of bipolar resistive switching (a) in air, (b) in vacuum and (c) in pure oxygen ($1 \times 10^4$ Pa). The upper halves and the bottom halves of the dotted boxes were written and erased with +5 V and -5 V, respectively. Then the current distributions were mapped with a +0.2 V read voltage. (d) Deposition-oxygen-pressure dependence of ratio of IRS/LRS for NiO films measured in atmosphere (blue square), and ambient-oxygen-pressure dependence of ratio of IRS/LRS for the 10 Pa-deposited NiO film during measurement (red circle). The ratios of IRS/LRS were calculated according to the CAFM current mappings, where the LRSs were stimulated with +5 V, and the current distributions were mapped with a +0.2 V read voltage. The two points on the shadow line were obtained from CAFM current mapping of (e) NiO film deposited at a 10 Pa oxygen pressure and measured at a 100 Pa oxygen pressure and (f) NiO film deposited at a 0.1 Pa oxygen pressure and measured under 1 atm. Scale bar: 200 nm.

The oxygen migration inside NiO film at positive/negative bias changes the stoichiometry in the region near the BE, inducing the LRS and HRS, respectively.[15] Therefore, the region near the BE should be recognized as the active switching region, and the elemental compositions therein should be different for NiO films with different memristive behaviors. In addition, the CAFM result revealed that the conducting filaments in the LRS distribute uniformly over the whole scanned area, including grains and grain boundaries (Figure S5), and the diameter of the conducting filaments is around 10 nm. Therefore, to simplify the measurement and analysis, we focused the characterization on the region inside the columnar grains, wherein the structural characterization on an atomic scale was carried out near the BE. Figure 4a shows the interface between NiO and Pt BE. The NiO films were (111)-oriented on the (111)-oriented Pt BE, which were also confirmed by the X-ray diffraction (XRD) results (Figure S10), demonstrating that NiO films grown at various oxygen pressures share the same crystal orientation.

STEM characterization was carried out in the region near the BE to identify oxygen vacancies. Figure 4b,c displays the cross-sectional high-angle annular-dark-field (HAADF) and ABF micrographs of the NiO films deposited at 10 Pa and $9\times10^{-4}$ Pa, respectively, viewed along the [1$\bar{1}$0] axis. Ni sites are clearly visible in the HAADF images while O sites are masked because of the HAADF contrast roughly proportional to the 1.7th power of the atomic number (Z).[31] Both Ni and O sites can be distinguished in the ABF images in that the contrast of ABF micrograph is proportional to $Z^{1/3}$.[43] In the line profiles of the ABF micrographs along the yellow bars, the deeper and shallower valleys are recognized as Ni and O atomic columns, respectively. The O sites are marked with red arrows and the red/blue disks display the stacking sequences of Ni-O. The oxygen vacancy concentrations are compared based on the absence of valleys at O-sites,[32] which are depicted as lighter blue disks in the Ni-O sequences. Herein, it is confirmed that the concentration of oxygen vacancy in the $9\times10^{-4}$ Pa-deposited NiO film is higher than that in the 10 Pa-deposited one. More ABF images obtained from different areas of the two samples are available in Figure S11. The phenomenon of low deposition oxygen pressure producing more oxygen vacancies has also been reported in other materials such as ZnO and LaAlO$_3$.[44,45] This conclusion is also supported by the higher ratio between the intensities of O-valley and Ni-valley in the 10 Pa-deposited NiO film than that in the $9\times10^{-4}$ Pa-deposited one, which is visually recognizable in the ABF profiles.

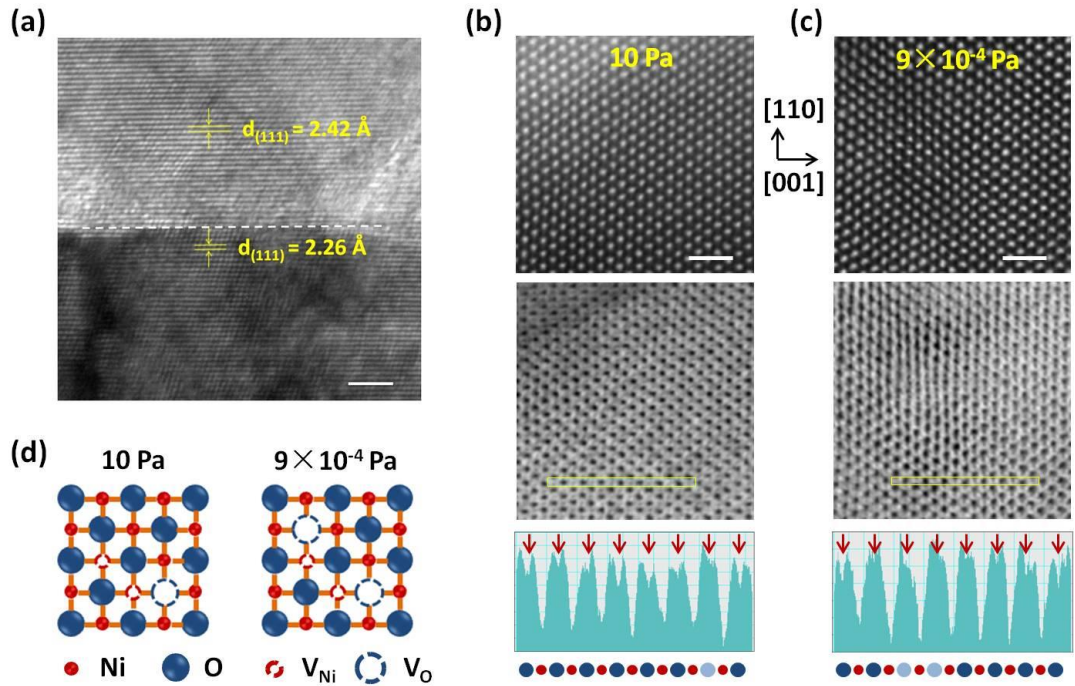

**Figure 4.** STEM observation of oxygen vacancies in the 10 Pa- and $9\times10^{-4}$ Pa-deposited NiO films. (a) TEM image of the interface between the 10 Pa-deposited NiO film and Pt BE. Below the dashed red line is Pt BE, above the dashed red line is NiO film, both are confirmed by measuring the interplanar spacings. Scale bar: 2 nm. STEM results of (b) 10 Pa-deposited NiO film and (c) $9\times10^{-4}$ Pa-deposited NiO film. The upper panels of both are HAADF images, the middle panels are ABF images and the bottom panels are line profiles along the yellow bars in ABF images. Scale bar: 1nm. Between (b) and (c) the in-plane orientations are indicated. (d) Schematic distributions of nickel and oxygen vacancies for the 10 Pa- and $9\times10^{-4}$ Pa-deposited NiO films, respectively. $V_{Ni}$ is nickel vacancy, and $V_O$ is oxygen vacancy.

The ABF STEM results provide information of oxygen vacancies, but the further information about nickel vacancies is hard to achieve here, so we employed EELS measurements to detect the mean valence state of Ni. In Figure 5, the O *K*-edge and Ni *L*-edge spectra for the 10 Pa- and $9\times10^{-4}$ Pa-deposited NiO films are displayed. Comparatively, the pre-peak in O *K*-edge spectra marked by the red arrow emerges for the 10 Pa-deposited NiO film, which is typical of nickel-deficient NiO.[34] Since there are both cation and anion vacancies coexisting in NiO, this result suggests that the 10 Pa-deposited NiO film has more nickel vacancies than oxygen vacancies, while the $9\times10^{-4}$ Pa-deposited NiO film has approximately equal concentrations of nickel and oxygen vacancies (undoped case or absence of $Ni^{3+}$), which is also validated by the comparison between ratios of $L_3/L_2$ in Ni $L$-edge spectra for the two samples. The ratio between the intensities of the $L_3$ and $L_2$ peaks can provide an estimation of Ni valence.[46] Higher valence of cation gives rise to a smaller ratio of $L_3/L_2$ due to the lower occupation of 3$d$-states.[47] It is discernable in Figure 5 that the ratio of $L_3/L_2$ for the 10 Pa-deposited NiO film is smaller than that for the $9\times10^{-4}$ Pa-deposited one, revealing that the former possesses higher Ni valence, which indicates the concentration surplus of Ni vacancy over O vacancy in the 10 Pa-deposited NiO film due to the charge neutrality requirement. It should be mentioned that a Ni vacancy and an O vacancy are prone to stay as the nearest neighbor due to its lowest formation energy as revealed by calculations.[29] Therefore, in the 10 Pa-deposited NiO film, each O vacancy is expected to be accompanied by a Ni vacancy, and the excessive Ni vacancies remain isolated, while in the $9\times10^{-4}$ Pa-deposited NiO film, the approximately equal amounts of nickel and oxygen vacancies are accompanied by each other. The schematic distributions of nickel and oxygen vacancies in the two samples are illustrated in Figure 4d. The concentration surplus of Ni vacancy over O vacancy is responsible for the large magnitude of bipolar memristive behaviors in the 10 Pa-deposited NiO film, which can be understood by the dual-defects-based model.

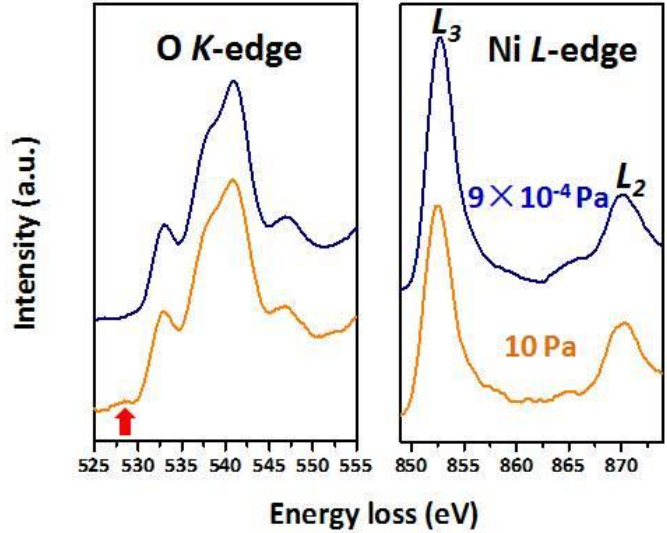

**Figure 5.** EELS results of the 10 Pa- and $9\times10^{-4}$ Pa-deposited NiO films. The O *K*-edge (left) and Ni *L*-edge (right) spectra for the 10 Pa- and $9\times10^{-4}$ Pa-deposited NiO films are shown. The pre-peak in O *K*-edge spectra of the 10 Pa-deposited NiO film is marked by the red arrow.

The mechanism of bipolar resistive switching of NiO is illustrated in details in Figure 6. For a high-oxygen-pressure deposited NiO film, the downward drift of oxygen ions and oxygen incorporation at the surface of NiO film occur simultaneously during the SET process (Figure 6a), the recombination of oxygen ions and oxygen vacancies results in a *p*-type conducting filament composed of isolated nickel vacancies. Due to the weaker electric field in the region near the BE than that around the CAFM tip, the formed conducting filament is consequently weaker in this region. During the RESET process (Figure 6b), the upward drift of oxygen ions breaks the conducting filament near the BE, which is assisted by the higher electric field across the region of weaker filament with higher resistivity. In the mean time, the upward oxygen migration compensates for the oxygen extraction in the upper region near the surface of NiO film. For the NiO films deposited at low oxygen pressures, the amount of oxygen vacancy is large. In this case, there are insufficient mobile oxygen ions to recombine the large amount of oxygen vacancies during the SET process, resulting in a weak or incomplete *p*-type conducting filament composed of low concentration of isolated nickel vacancies (Figure 6c), hence the magnitude of resistive switching is small. And when the memristive behaviors is measured in ambient with low oxygen pressure or in vacuum, there is no sufficient oxygen incorporation to compensate for the downward-drifted oxygen ions during the SET process. As a result, the formed conducting path is disconnected with low concentration of isolated nickel vacancies near the surface of NiO film, and the switching to LRS is hence interrupted (Figure 6d).

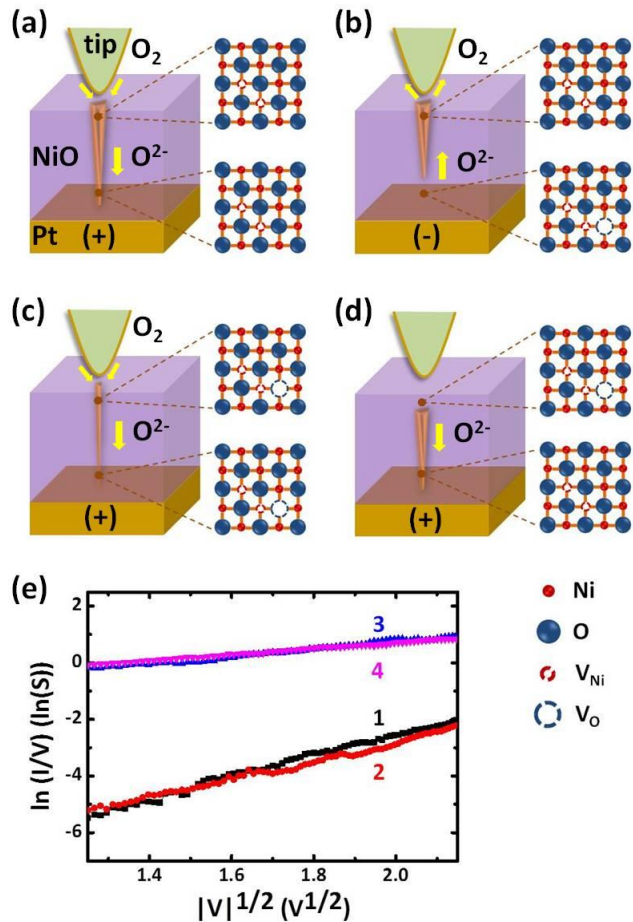

**Figure 6.** Mechanisms of bipolar resistive switching and electrical conduction of NiO. Schematic of (a) SET process and (b) RESET process in the bipolar resistive switching of high-oxygen-pressure deposited NiO film measured in ambient oxygen. The high-oxygen-deposited sample has isolated nickel vacancies, the concentration of which becomes even higher throughout the formed filament during the SET process, and lowers down in the region near the BE during the RESET process resulting in disconnection of the filament. (c) SET process for the low-oxygen-pressure deposited NiO film measured in ambient oxygen. The low-oxygen-pressure deposited sample has less or no isolated nickel vacancies, so the concentration of isolated nickel

vacancies in the filament is consequently lower. (d) SET process for the high-oxygen-pressure deposited NiO film measured at low ambient-oxygen pressure or in vacuum. Due to the insufficient oxygen incorporation from the ambient, the concentration of isolated nickel vacancies beneath the surface will not be increased during the SET process, so the filament disconnects near the surface. (e) Poole-Frenkel emission fitting of *I-V* curves for the LRS and HRS of the 10 Pa-deposited NiO film. The *I-V* curve was obtained by sweeping voltage in the sequence of -10 to 0 V, 0 to 10 V, 10 to 0 V and 0 to -10 V, which are labeled by numbers 1, 2, 3 and 4, respectively.

To explore the nature of electrical transport for the HRS and LRS, we carried out fitting to their *I-V* characteristics and found that Poole-Frenkel conduction mechanism can exclusively give a good fitting as shown in Figure 6e, and two other mechanisms, including Fowler-Nordheim tunneling and space charge limited current (SCLC) have been excluded as shown in Figure S12 and S13. The fitting ranges are between 1.6 V and 4.6 V, and between -4.6 V and -1.6 V, respectively. Poole-Frenkel conduction mechanism belongs to the bulk-limited case and the current density can be described by the following equation[48]

$$J \propto E \exp\left[\frac{-q\left(\phi_B - \sqrt{qE/\pi\varepsilon_0\varepsilon_r}\right)}{kT}\right]. \qquad (1)$$

*J* is the current density, *q* is the electronic charge, $\varepsilon_0$ is the permittivity of free space, $\varepsilon_r$ is the high-frequency dielectric constant, *E* is the electric field, and *k* is the Boltzmann's constant, $\phi_B$ is voltage barrier in zero electric field. In the $\ln(I/V) \sim V^{1/2}$ plot, the slope of the fitting line 1 reveals that the dielectric constant of the HRS is 12.3, quite close to the high-frequency value of NiO,[49] while the dielectric constant of LRS obtained from the slope of fitting line 3 is 102.7. Since a conducting filament forms in the NiO matrix in the LRS, the large dielectric constant of LRS can be understood with the percolation threshold transition, in which the dielectric constant can be increased with orders of magnitude.[50] It should be mentioned that the oxygen ion migration should not have obvious impact on the electric conduction in the fitting range due to the relatively low voltage and short biasing time.[51,52] Therefore, electrical transport for both the HRS and LRS can be described by Poole-Frenkel conduction mechanism and suggests that holes introduced by Ni vacancies are trapped.[53]

Interestingly, it has been proposed that the metal interstitials instead of oxygen vacancies, can act equivalently well in the memristive behaviors of many metal oxides.[54] However, we found that Ni interstitial is energetically unfavorable in NiO, according to our first-principles calculations of formation energies for different native point defects (Figure S14). Therefore, the metal interstitials should not be expected to play a significant role in the memristive behaviors of NiO. On the other hand, the defect chemistry of NiO has been well known from the mid-sixties of the last century that only nickel vacancies contribute to the electrical properties of NiO.[55-58] However, if the work temperature and oxygen pressure are relatively low, the oxygen vacancy is expected to play an important role.[59] The oxygen vacancies observed by STEM in our samples, which were prepared at 500 ℃ and at oxygen pressures ranging from 10 Pa to $9\times10^{-4}$ Pa, and their contributions to the memristive behaviors in dual-defects-based model are quite reasonable.

## 4. CONCLUSIONS

In summary, CAFM measurement shows typical bipolar memristive behaviors and good device performance of NiO films. By choosing the oxygen pressures during film deposition and CAFM measurement, the occurrence and magnitude of bipolar memristive behaviors of NiO can be controlled. Based on the structural and chemical analysis for NiO films with different memristive behaviors by STEM and EELS, we demonstrate that the concentration surplus of Ni vacancy over O vacancy determines the memristive behaviors. The inner and ambient oxygen work collaboratively to facilitate memristive behaviors, involving the compensatory processes of inner-oxygen-ions drift and surface-oxygen exchange. Our work conforms to the dual-defects-based model, which describes the modulation of concentration of isolated nickel vacancies by oxygen ions migration, and reveal the important role of cation vacancy in the memristive behaviors. It also provides a methodology to investigate the effect of dual defects on memristive behaviors.

**ASSOCIATED CONTENT**

**AUTHOR INFORMATION**

**Corresponding Author**

*(Y.Z.) E-mail: ygzhao@tsinghua.edu.cn.

**Author Contributions**

Z.S. and Y. Z. conceived the idea, designed the experiments and interpreted the results. Z.S. and D. Z. performed the device fabrications. Z.S. carried out the electrical measurements and XRD experiments. M.H., L.G., Q. Z. and C.-W.N. performed the TEM, STEM and EELS characterizations and analyzed the results. C.M. and K. J. conducted the theoretical fittings. N. L., N.W. and W. D. did the first-principles calculations. Z.S. and Y. Z. prepared the manuscript. All authors discussed the results and implications and commented on the manuscript at all stages.

**Notes**

The authors declare no competing financial interest.

**ACKNOWLEDGMENTS**

We gratefully acknowledge Y. Wang, X. Shi and X. Lai for help with the CAFM measurements. This work was supported by the National Science Foundation of China (Grant No. 11134007 and 51332001), National Basic Research Program of China (Grant Nos. 2015CB921402,

2015CB921400, 2012CB921702 and 2014CB921002), Special Fund of Tsinghua for basic research (Grant No. 201110810625), and Strategic Priority Research Program of the Chinese Academy of Sciences (Grant No. XDB07030200).